\documentclass[12pt,preprint]{aastex}

\usepackage{amsmath}

\shorttitle{Limit cycles can reduce the width the habitable zone}
\shortauthors{Haqq-Misra et al.}

\begin{document}


\title{Limit cycles can reduce the width of the habitable zone}


\author{Jacob Haqq-Misra\altaffilmark{1,2}, 
Ravi Kumar Kopparapu\altaffilmark{1,2,3,4}, 
Natasha E. Batalha\altaffilmark{2,5,6},  
Chester E. Harman\altaffilmark{2,6,7}, and
James F. Kasting\altaffilmark{2,6,7}}


\altaffiltext{1}{Blue Marble Space Institute of Science, 1001 4th Ave Suite 3201, Seattle, WA 98154, USA}
\altaffiltext{2}{NASA Astrobiology Institute's Virtual Planetary Laboratory, P.O. Box 351580, Seattle, WA 98195, USA}
\altaffiltext{3}{NASA Goddard Space Flight Center, 8800 Greenbelt Road, Mail Stop 699.0 Building 34, Greenbelt, MD 20771, USA}
\altaffiltext{4}{Department of Astronomy, University of Maryland, College Park, MD 20742, USA}
\altaffiltext{5}{Department of Astronomy and Astrophysics, The Pennsylvania State University, University Park, PA 16802, USA}
\altaffiltext{6}{Center for Exoplanets and Habitable Worlds, The Pennsylvania State University, University Park, PA 16802, USA}
\altaffiltext{7}{Department of Geosciences, The Pennsylvania State University, University Park, PA 16802, USA}


\begin{abstract}
The liquid water habitable zone (HZ) describes the orbital distance at which a 
terrestrial planet can maintain above-freezing conditions through regulation by the carbonate-silicate cycle.
Recent calculations have suggested 
that planets in the outer regions of the habitable zone cannot maintain stable, warm climates, 
but rather should oscillate between long, globally glaciated states and shorter periods of 
climatic warmth. Such conditions, similar to `Snowball Earth' episodes experienced on Earth, 
would be inimical to the development of complex land life, including intelligent life. Here, 
we build upon previous studies with an updated an energy balance climate model to 
calculate this `limit cycle' region of the habitable zone where such cycling would occur. 
We argue that an abiotic Earth would have a greater CO$_2$ partial pressure than today because 
plants and other biota help to enhance the storage of CO$_2$ in soil. 
When we tune our abiotic model accordingly, we find that limit cycles 
can occur but that previous calculations have overestimated their importance. 
For G stars like the Sun, limit cycles occur only for planets with CO$_2$ 
outgassing rates less than that on modern Earth. For K and M star planets, 
limit cycles should not occur; however, M-star planets may be inhospitable to life for other reasons.  
Planets orbiting late G-type and early K-type stars retain the greatest potential
for maintaining warm, stable conditions. 
Our results suggest that host star type, planetary volcanic activity, and seafloor weathering
are all important factors in determining whether planets 
will be prone to limit cycling.
\end{abstract}



\section{Introduction}

Earth's orbit falls within the boundaries of the habitable zone (HZ) where a rocky planet 
can maintain liquid water on its surface, 
given a CO$_2$-N$_2$-H$_2$O atmosphere and some mechanism (e.g., plate tectonics) for recycling these volatiles
\citep{kasting1993,abe2011,kopparapu2013,kopparapu2014,leconte2013,wolf2014,yang2014}.
The inner HZ edge is bounded by either the 
runaway greenhouse effect, in which liquid water evaporates entirely, or the 
moist greenhouse\footnote{We use the term `moist greenhouse' throughout our paper, although 
we acknowledge that this may be imprecise terminology, as Earth’s present atmosphere could be described as a moist greenhouse. Other terms such as 
`moist stratosphere' or `diffuse tropopause' might be better descriptors of this phenomenon, but they suffer from sounding arcane or obscure.}
effect, in which liquid water persists on a planet's surface but the stratosphere becomes 
wet and water is lost by photodissociation followed by escape of hydrogen to space. The 
moist greenhouse effect is difficult to simulate with one-dimensional models that assume 
a saturated troposphere \citep{kopparapu2013}, but it does appear in some general circulation models \citep{wolf2014}. 
Planets farther out in the HZ are expected to accumulate dense CO$_2$ atmospheres because of 
the negative feedback between silicate weathering (the loss process for atmospheric CO$_2$) 
and surface temperature \citep{walker1981}. 
But this feedback loop is ultimately limited by CO$_2$ 
condensation and Rayleigh scattering, which combine to create an outer HZ boundary 
termed the `maximum greenhouse limit' \citep{kasting1993}. 
A proposed extension of the outer HZ boundary by formation of CO$_2$ ice clouds \citep{forget1997,mischna2000,colaprete2003,forget2013}
appears less likely when the `scattering greenhouse effect' of these clouds is recomputed using more accurate radiative transfer models, 
\citep {kitzmann2016}.

The conventional thinking regarding the HZ outer edge may be too optimistic, however, because 
it fails to account for mass transfer rates of CO$_2$. CO$_2$ is released from volcanoes and is 
consumed by silicate weathering followed by deposition of carbonate sediments \citep{walker1981,berner1983}. These 
processes are in approximate balance on modern Earth, creating a climate that is stable and 
relatively warm, even if it is sometimes perturbed by glacial-interglacial cycles. Occasional 
`Snowball Earth' episodes in which the planet is fully glaciated \citep{hoffman1998} have been attributed to 
a variety of complicating factors, including changes in atmospheric O$_2$ and CH$_4$ \citep{pavlov2003}, as well 
as limit cycles involving atmospheric CO$_2$ \citep{tajika2007,mills2011}. Limit cycles---oscillations between ice-free 
and globally glaciated states---occur in models of the early Earth in which volcanic outgassing 
rates are too low to sustain a CO$_2$-warmed climate \citep{tajika2007}.

Several new papers have argued that such limit cycle behavior could be far more prevalent than 
previously thought \citep{kadoya2014,kadoya2015,menou2015}. Depending on the volcanic outgassing rate
and host stellar type, 
some Earth-like planets are 
subject to climatic limit cycles when stellar insolation is low, as it is in the outer parts 
of a star's HZ. All of these recent studies employed parameterized versions of 
energy-balance climate models (EBMs). Such models include radiative balance between incident 
stellar and outgoing infrared radiation, along with diffusional heat fluxes between different 
latitude bands. These models are useful tools for studying climate variations that occur on 
time scales of thousands to tens of millions of years.

Here we discuss the possibility that limit cycles could reduce the outer edge of the HZ.
We use our own EBM, which implements an updated parameterization of radiative transfer 
based on 1-D radiative-convective HZ calculations \citep{kopparapu2013,kopparapu2014}.
 Our model includes a representation 
of the carbonate-silicate cycle used by previous studies \citep{menou2015}, and also expands 
on previous work to include the effect of CO$_2$ condensation and the impact of seafloor weathering.
These improvements allow us to determine, for the first time, the limit cycle boundaries relative to 
the conventional liquid water HZ for different types of stars.

\section{Model Description}

Latitudinal energy balance models (EBMs) are computationally efficient models that are well-suited 
to exploring glacial cycling and its influence on climate. Although EBMs can be useful 
with a highly idealized linearization of infrared absorption \citep{north1981,gaidos2004,haqqmisra2014},
investigation of HZ limits requires an EBM with a more sophisticated radiative transfer 
parameterization \citep{williams1997,fairen2012,vladilo2013,kadoya2014,menou2015}.
However, no previously published EBM has yet been updated to include parameterizations based upon 
radiative-convective HZ model calculations \citep{kopparapu2013,kopparapu2014}, which 
use coefficients derived from the HITRAN 2008 \citep{rothman2009} 
and HITEMP 2010 \citep{rothman2010} spectroscopic databases to
account for additional 
absorption features of H$_2$O and CO$_2$ compared to earlier 
models \citep{kasting1993}. Existing EBMs that consider planetary 
habitability \citep{vladilo2013,kadoya2014,kadoya2015,menou2015} use an older radiation scheme \citep{williams1997}, 
which is based upon a polynomial parameterization of prior radiative-convective calculations 
\citep{kasting1991}. These EBM investigations continue 
to be useful for interpreting exoplanet habitability and guiding more complicated modeling studies 
toward interesting regions of parameter space, but the lack of up-to-date radiative transfer parameterization 
in such models can limit their utility when comparing with other recent climate models or interpreting exoplanet observations. 

We use a one-dimensional EBM that has been developed in 
previous studies \citep{williams1997,gaidos2004,fairen2012,haqqmisra2014}. This EBM calculates 
meridionally averaged temperature $T$ as a function of latitude $\theta$ and time $t$ according 
to the equation
\begin{equation}
C\frac{\partial T}{\partial t}=\bar{S}\left(1-\alpha\right)-F_{OLR}+\frac{1}{\cos\theta}\frac{\partial}{\partial\theta}\left(D\cos\theta\frac{\partial T}{\partial\theta}\right)\label{eq:EBM}.
\end{equation}
Eq. (\ref{eq:EBM}) expresses the change in temperature as the sum of stellar heating, infrared cooling, and meridional diffusion. 
Diurnally averaged solar flux $\bar{S}=S \cdot q(\theta)$ is the product of a constant solar flux $S$ and a 
function of latitude $q(\theta)$ (which assumes a circular orbit) to yield seasonally varying 
insolation \citep{gaidos2004}. The effective heat capacity of the surface and atmosphere, $C$,  
depends on the fraction of ocean and ice coverage at a given latitude 
\citep{williams1997,fairen2012}. We prescribe Earth-like 
conditions by setting 70\% ocean coverage at all latitudes and allowing fractional ice coverage between 263 K to 
273 K. Radiative fluxes are represented by the top-of-atmosphere albedo $\alpha$ and the 
infrared outgoing radiative flux $F_{OLR}$. 
Our FORTRAN implementation of 
this EBM is discretized into 18 equally spaced latitudinal zones with a initial uniform temperature profile of 
$T = 233\mbox{ K}$ and stepped through 1000 (or more) complete orbits by increments of $\Delta t=8.64\times10^{3}\mbox{ s}$
to numerically solve Eq. (\ref{eq:EBM}) 
with a forward finite differencing scheme\footnote{We can be confident that our choice 
of parameters will yield a converged solution by examining the CFL condition for numerical 
stability, $C \leq C_{max}$, where $C_{max} = 1$ for explicit time-marching solvers like ours.
If we approximate our length interval as the radius of Earth divided by the total 
number of latitudinal bands, and if we assume a typical advective wind speed on Earth of 
$u = 10$ m s$^{-1}$, then $C = u\Delta t/\Delta x = 0.24 < C_{max}$. Only when the number 
of latitude bands approaches $\sim$75 does $C \approx C_{max}$.}. 
We initialize all simulations with snowball conditions 
($T = 233\mbox{ K}$) and CO$_2$ partial pressure $p$CO$_2 = 3.3\times10^{-4}\mbox{ bar}$.

The diffusive parameter, $D$, 
describes the efficiency of meridional energy transport and scales with changes in 
atmospheric pressure, heat capacity, atmospheric mass, and rotation rate \citep{williams1997}.
This parameter accounts for the exchange of sensible and latent energy fluxes 
between the tropics and midlatitudes by adjusting the energy balance equilibrium 
temperature at each latitude.
\citet{williams1997} assumed that 
$D$ is proportional to the inverse square of rotation rate, so that a more rapidly (slowly) rotating 
planet will show a decreased (increased) tendency toward meridional energy distribution. This 
behavior mimics well-known results from general circulation models \citep{williams1982,showman2013}
while still maintaining the computational efficiency of an EBM. 
We focus our study on a model planet with a rotation rate equal to present-day Earth, but this 
assumption does not significantly alter our results. 
A more slowly rotating planet would have increased meridional energy transport and would therefore 
tend toward uniform conditions at all latitudes; such a limit is analogous to the globally-averaged EBM used 
by \citet{menou2015}. A more rapidly rotating planet would show more extreme variations between 
equatorial and polar temperatures, but this contrast does not significantly alter 
the onset of global glaciation and deglaciation events. Sensitivity studies 
indicate that either of these cases will still 
exhibit transitions into a limit cycle state at approximately the same value of solar constant, so our assumption 
of a fixed present-day rotation rate does not necessarily limit the scope of our calculations.

Surface albedo $a_s$ is calculated as a weighted sum of the albedos of unfrozen land, 
unfrozen ocean, and ice coverage at each latitude. We assume an albedo of 0.2 for 
unfrozen land, while the albedo of ocean varies as a function of solar zenith angle 
\citep{williams1997,fairen2012}. We also implement a band-dependent ice albedo \citep{pollard2005} 
that partitions the broadband ice albedo $a_i$ into a visible component, $a_{i,vis} = 0.8$, and a 
near-infrared component, $a_{i,nir} = 0.5$. This latter value applies to snow-covered ice, 
as we discuss further below. According to 
\citet{shields2013}, the near-IR albedo of bare ice is $\sim$0.3. Our model 
does not include a hydrologic cycle, and so it does not calculate the ratio 
of bare ice to snow-covered ice, as previous studies \citep{shields2013} did. 
For a given stellar spetrum, we define total ice albedo 
as the sum $a_i = a_{i,vis}f_{vis} + a_{i,nir}f_{nir}$, where $f_{vis}$ and $f_{nir}$ are the respective percent 
contributions of visible ($\le$ 700 nm) and near-infrared ($\textgreater$ 700 nm) radiation. 
Following a previously published method \citep{kopparapu2013,kopparapu2014}, we use “BT-Settl” model spectra 
\citep{allard2007} to calculate the percent contribution of visible and near-infrared radiation 
for F-type (7200 K, 67\% visible), G-type (5800 K, 52\% visible), K-type (4600 K, 32\% visible), 
and M-type (3400 K, 10\% visible) stars.  

We parameterize the infrared outgoing radiative flux $F_{OLR}$ as a fourth-order polynomial function 
of the partial pressure of carbon dioxide, $p$CO$_2$, and surface temperature, $T$. Likewise, we parameterize 
top-of-atmosphere albedo $\alpha$ as a third-order polynomial function dependent on $p$CO$_2$, $T$, surface albedo 
$a_s$, and stellar zenith angle. Assuming a noncondensible background pressure of 1 bar N$_2$, we use a
radiative-convective climate model developed for HZ calculations \citep{kopparapu2013,kopparapu2014} to obtain best fits of more than 50,000 
calculations for $F_{OLR}$ and $\alpha$ over a parameter space spanning $10^{-5}\mbox{ bar} < p\mbox{CO}_2 < 35\mbox{ bar}$, 
$150\mbox{ K} < T < 350\mbox{ K}$, and $0.2 < a_s < 1$, across all zenith angles. We calculate separate radiative 
transfer parameterizations for F, G, K, and M stars, using model spectra \citep{allard2007}, which we 
describe further
in the Appendix.

We assume in our EBM that any condensing CO$_2$ accumulates on the surface as dry ice. CO$_2$ 
condensation occurs when $p$CO$_2$ exceeds the CO$_2$ saturation vapor 
pressure at surface temperature $T$ within a latitudinal zone. We assume that dry ice will radiatively dominate over
water ice, so we set total ice albedo $a_i = 0.35$ for frozen carbon dioxide \citep{warren1990} when CO$_2$ condensation 
occurs. We also keep an inventory of the thickness of CO$_2$ ice that condenses or melts on the surface at a given latitude, 
and we adjust the radiative contribution of CO$_2$ at each latitude, as well as the global value of $p$CO$_2$, by 
a corresponding amount each iteration. We calculate the thickness $z$ of accumulating ice as 
$z=\Delta (p\mbox{CO}_{2})/g\rho$, where $\Delta (p\mbox{CO}_{2})$ is the partial pressure of CO$_2$ that condenses 
into ice, $g = 9.81\mbox{ m s}^{-2}$, and $\rho = 1600\mbox{ kg m}^3$ is the density of dry ice. Assuming 
that ice thickness is limited only by geothermal heat flow, we express the maximum CO$_2$ ice 
thickness as: $z_{max} = k\Delta T/F_g$ \citep{pollard2005}, where 
$k = 0.6\mbox{ W m}^{-1}\mbox{ K}^{-1}$ is the thermal conductivity of solid CO$_2$ 
\citep{kravchenko1986,stewart2002}, $\Delta T$ is the 
temperature difference between the atmosphere and seawater beneath the ice, and $F_g = 0.1\mbox{ W m}^{-2}$ 
is an Earth-like geothermal heath flux. For a temperature difference $\Delta T = 25\mbox{ K}$ characteristic of a globally 
glaciated planet \citep{pollard2005}, this gives a maximum CO$_2$ ice thickness of $z_{max} = 150\mbox{ m}$. Any 
additional accumulation would result in basal melting of CO$_2$ glaciers and the transport 
of liquid CO$_2$ to lower latitudes, although 
none of our simulations in this study reach conditions where $z > z_{max}$.

Our latitudinal model can only represent clouds through adjustments to surface albedo, 
and we assume in our calculations that water clouds cover half of the surface. We have 
explored the sensitivity of climate to this cloud fraction parameter in a previous study \citep{fairen2012},
which showed that excessive cloud cover can cause a planet to plummet into global glaciation. 
We account for the absorption of infrared radiation by clouds by subtracting a fixed amount 
of 8.5 W m$^{-2}$ from $F_{OLR}$ at each latitude band, following \citet{williams1997}. 
This value was selected by requiring 
that the EBM should produce a present-day Earth temperature of 288 K when the model is initialized 
with above-freezing initial conditions at $S/S_0 = 1.0$.
Possible sources of additional warming include CO$_2$ ice clouds, which could warm the surface
by up to 15 K by providing additional downward-directed 
infrared radiation through a scattering greenhouse effect 
\citep{forget1997,mischna2000,colaprete2003,forget2013}. 
But this is not by itself enough to 
deglaciate a planet \citep{forget2013}, and \citet{kitzmann2016} has shown that these previous calculations may have 
significantly overestimated the warming from such clouds.
By neglecting their radiative impact, we generate a somewhat 
pessimistic outer limit on planetary habitability.

\subsection{Outgassing Rates}

CO$_2$ accumulates in our model atmosphere as a result of the carbonate-silicate cycle, 
which allows a frozen planet to eventually deglaciate.
We represent the $\sim$0.5 Myr timescale of the carbonate-silicate cycle 
with the time variable, $\tau$, to contrast this slower timescale from the faster time step, $t$, used in 
Eq. (\ref{eq:EBM}). Following \citet{menou2015} we implement the time evolution of $p$CO$_2$ into our EBM according to
\begin{equation}
\frac{d}{d\tau}(pCO_{2})=V-W-W_{sea}\label{eq:co2evo},
\end{equation}
where $V$ represents the volcanic outgassing rate of CO$_2$, $W$ represents the uptake of CO$_2$ by rock weathering, 
and $W_{sea}$ represents the uptake of CO$_2$ by seafloor weathering. We begin by assuming volcanic outgassing 
to be a constant $V=V_{\Earth}$, where $V_{\Earth}$ is the present-day value. 

We also re-examine prior assumptions regarding the total rate of 
CO$_2$ outgassing.
\citet{menou2015} 
assumed a value of $V_{\Earth}=7\mbox{ bar/Gyr}$, which, converted to geochemists' units for an Earth-mass planet, 
corresponds to rate of $V_{\Earth}=0.83\mbox{ Tmol/yr}$. 
\citet{menou2015} appears to have
obtained this value from \citet{abe2011}, who in turn obtained this value from \citet{saal2002} 
as the expected CO$_2$ flux from the depleted mantle alone.
However, best estimates of the 
total terrestrial CO$_2$ outgassing rate
suggest $V_{\Earth}\approx7.5\mbox{ Tmol/yr}$ \citep{gerlach2011,jarrard2003}, which is about a factor of ten 
greater than the value used 
by \citet{menou2015}. By comparison, \citet{tajika2007} assumed a modern outgassing rate of 8 Tmol/yr.
We argue that this modern rate is the best choice for assessing the habitability of 
a planet with Earth-like tectonic activity. 
We use a value of $V_{\Earth}=70\mbox{ bar/Gyr}$ in most of our calculations, 
but we also
explore 
the dependence of $V$ on defining the limit cycle HZ boundary.

\subsection{Weathering Rates}

The weathering rate in our model includes both land and seafloor processes.
For abiotic conditions, \citet{menou2015} 
adopted a functional form of $W$ as \citep{berner2001}
\begin{equation}
\frac{W}{W_{\Earth}}=\left(\frac{pCO_{2}}{p_{\Earth}}\right)^{\beta}e^{k_{act}(T-288)}\left[1+k_{run}(T-288)\right]^{0.65}\label{eq:weathering},
\end{equation}
where $W=W_{\Earth}\equiv V_{\Earth}$ for $T_{surf}=288\mbox{ K}$, $k_{act} = 0.09$ is an activation
energy, and $k_{run} = 0.045$ is a runoff efficiency factor. 
The value of $\beta$ is a critical factor 
determining the rate of $p$CO$_2$ evolution and has been estimated to be in the range of 0.25 to 1.0 
in the absence of vascular plants \citep{kump2000,berner2001,abbot2012}.
A lower value of $\beta = 0.25$ is appropriate for
environmental conditions with pH $<$ 5, and probably represents a minimum dependence
of weathering rate on $p$CO$_2$ \citep{berner1992}. 
\citet{tajika2007} assumed $\beta = 0.3$, following \citet{walker1981}.
Their parameterization, in turn, was based on laboratory studies of silicate dissolution by 
\citet{lagache1976}. It remains unclear 
whether or not the presence of widespread biological systems should affect $\beta$, although 
the presence of life should increase the weathering rate $W$ \citep{cawley1969,schwartzman1989,berner1992}.
We use a value of $\beta$ = 0.5 for most of our calculations, 
which is appropriate for conditions under which the weathering rate is proportional to dissolved [H$^+$].
We also perform a more limited set of calculations for $\beta = 0.3$.

Seafloor weathering has been neglected in previous calculations of limit cycling, 
but that is probably an oversight, particularly for planets with smaller amounts of land area than Earth.
We consider the effects of seafloor weathering by assuming that the seafloor weathering rate is 
independent of temperature according to the functional form
\begin{equation}
W_{sea}=W_{s}\left(\frac{pCO_{2}}{p_{\Earth}}\right)^{\gamma}\label{eq:seaweather},
\end{equation}
where $\gamma$ is a seafloor weathering parameter analogous to $\beta$ in Eq. (\ref{eq:weathering}) 
and $W_s$ is the baseline seafloor weathering that occurs in the absence of any $p$CO$_2$ dependence. 
Increases in the value of $W_{sea}$ or $\gamma$ cause a reduction in the frequency of limit cycling 
until a state of permanent glaciation occurs. This parameter $\gamma$ may be near unity, but weaker 
weathering dependence with $\gamma = 0.4$ or lower may also be consistent with some computed 
histories of early Earth \citep{sleep2001}. We initially assume $W_s = 0$ in our calculations to assess the limit 
cycle boundary of the HZ, and we then consider sensitivity to changes in the seafloor weathering rate.

\subsection{Partial pressure of CO$_2$ in soil}

A critical factor in determining the onset of limit cycles in the HZ 
is the partial pressure of CO$_2$ in soil.
The parameter $p_{\Earth}$ represents the long-term balance between atmospheric and soil CO$_2$ for 
present-day Earth. \citet{menou2015} assumed a value of $p_{\Earth} = 3.3\times10^{-4}\mbox{ bar}$ 
to represent the pre-industrial CO$_2$ level. However, the long-term balance between weathering 
and volcanism should be based on the value of $p$CO$_2$ in soil, rather than in the atmosphere. 
This implies that an abiotic Earth should have a higher value of atmospheric 
pCO$_2$ than today. To put it another way, if all of life were to suddenly vanish, then 
pCO$_2$ should increase until the atmospheric and soil (regolith) partiial pressures were equal.
If we wish to use our model to test the habitability of abiotic planets, or of inhabited planets that lack vascular land plants, 
then we should tune the model to an abiotic state.

For a biotic planet like present-day Earth, root respiration by vascular plants 
increases the value of soil $p$CO$_2$ by a factor of 10 to 100 \citep{kump2010}. 
We assume here that the enhancement is a factor of 30, in which case soil $p$CO$_2$ should be 
approximately $10^{-2}$ bar. For $\beta = 0.5$, this implies that land plants accelerate 
silicate weathering by a factor of $30^{0.5} \approx 5.5$. For $\beta = 0.3$, 
the acceleration would be $30^{0.3} \approx 2.8$. In either case, 
an abiotic present-day Earth would be warmer than today because land plants would no longer be 
pumping atmospheric CO$_2$ into soil. We therefore choose a value of $p_{\Earth} = 10^{-2}\mbox{ bar}$ 
for our calculations of the limit cycle HZ boundary.

\section{Results}

\begin{figure}
\epsscale{0.8}
\plotone{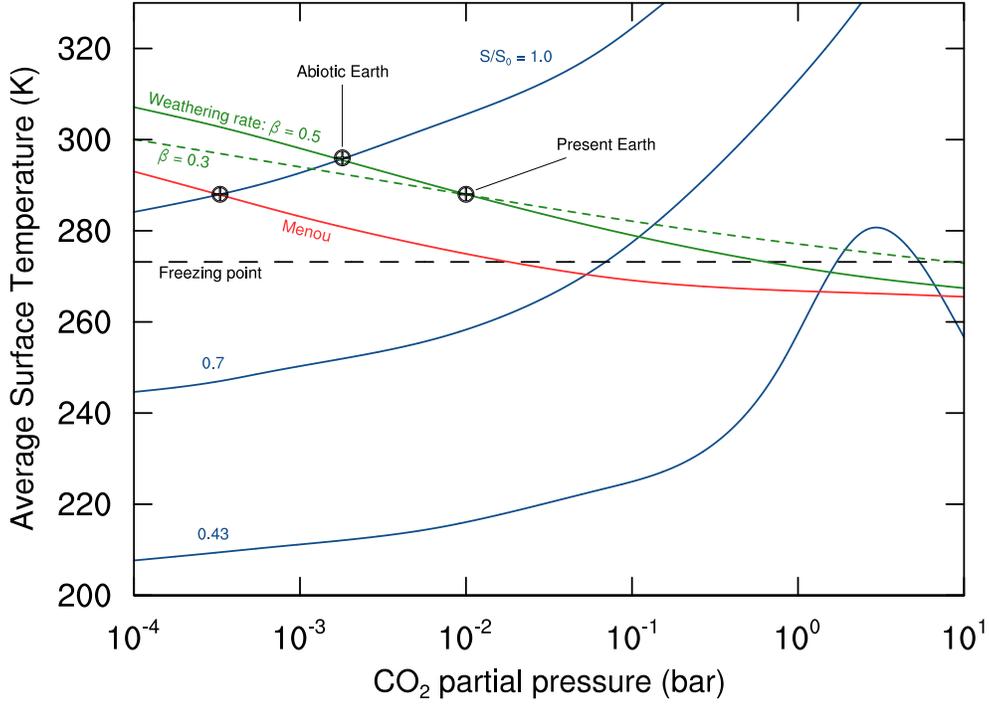}
\caption{Climate calculations indicate where limit cycling should occur. Average surface 
temperature as a function of $p$CO$_2$ (blue curves) is shown for $S/S_0$ = 1.0 
(present-day Earth), 0.7 (early Earth), and 0.43 (present-day Mars). 
Weathering rate curves indicate a model tuned to atmospheric $p$CO$_2$ = $3.3\times10^{-4}$ bar as assumed 
by \citep{menou2015} (red curve) and soil $p$CO$_2$ = $10^{-2}$ bar as we argue in this study (green curves).
Warm stable solutions occur where climate calculations intersect the weathering rate curve above the 
freezing point, when outgassing and weathering of CO$_2$ reach a steady-state. For the present-day value 
of solar flux ($S/S_0 = 1.0$), this intersection indicates the average surface temperature for an abiotic 
Earth. Note that our estimate of the abiotic Earth temperature is higher than predicted by \citet{menou2015} 
because we have tuned our model to present-day soil $p$CO$_2$.
Limit cycling occurs where climate calculations intersect the weathering rate 
below the freezing point, when CO$_2$ accumulates until deglaciation and the onset of weathering. 
\label{fig:cycles}}
\end{figure}

We first consider an Earth-like (but abiotic) planet orbiting a G-type star like the Sun.
At present-day stellar flux ($S/S_0 = 1.0$), our weathering model for abiotic Earth with $\beta = 0.5$ balances at 
soil $p$CO$_2$ = $1.8\times10^{-3}$ bar and average surface temperature of 296 K, while present Earth (with life) 
at a temperature of 288 K has a higher value of soil $p$CO$_2$ = $10^{-2}$ bar (Fig. \ref{fig:cycles}). By contrast, 
\citet{menou2015} argued that the carbon cycle on an abiotic Earth 
should balance at 288 K and $p$CO$_2$ = $3.3\times10^{-4}$ bar. 
This would only be true if land plants had zero effect on silicate weathering.  
Our model agrees with \citet{menou2015} in predicting no limit cycles for present-day Earth, but 
our results differ toward the outer edge of the HZ. At $S/S_0 = 0.7$, our model predicts stable 
warm climates above the freezing point, whereas \citet{menou2015} predicts limit cycles with 
prolonged glacial conditions.
Further outward in the HZ at $S/S_0 = 0.43$ (the effective solar flux at Mars' orbit), 
our model and \citet{menou2015} both predict that the intersection between the 
weathering rate curve and the greenhouse effect curve
falls beneath the freezing point, suggesting that limit cycles should occur. 
However, when we assume a value of $\beta = 0.3$ in our model, limit cycles do not occur at all (Fig. \ref{fig:cycles}, dashed green curve).
Elsewhere \citep{batalha2016}, we have argued that limit cycles are even more 
likely to have occurred on early Mars when the solar flux was significantly 
lower than today. For early Mars, though, the greenhouse effect must be supplemented 
by some absorber other than H$_2$O or CO$_2$; otherwise, even brief recoveries to 
above-freezing surface temperatures are impossible without outside stimuli, e.g., impact events \citep{segura2002}.

\begin{figure}
\epsscale{0.8}
\plotone{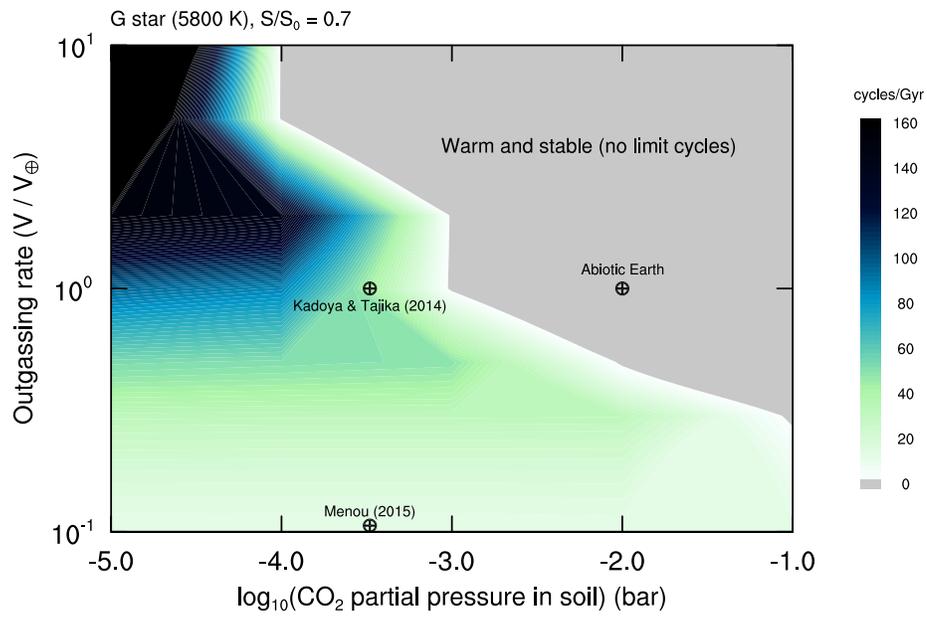}
\caption{Limit cycling is a function of outgassing rate and the partial pressure of CO$_2$ in 
soil. The limit cycle frequency is shown for a G star at $S/S_0 = 0.7$, with the grey region illustrating
the region of parameter space where limit cycles do not occur. The point labeled `Abiotic Earth' 
represents our preferred choice of parameters for uninhabited terrestrial planets.
\label{fig:soil}}
\end{figure}

The primary difference between the results of our model and those of \citet{menou2015} 
is caused by our higher assumed volcanic outgassing rate and by our treatment of the CO$_2$
partial pressure of in soil. The relevance of these factors 
in determining the frequency of limit cycle events is shown in Fig. \ref{fig:soil}, which shows 
the limit cycle frequency (in units of cycles per Gyr) as a function of $V$ and $p_{\Earth}$.
The grey region of this figure represents warm, stable solutions that are not prone to limit 
cycling. When we select $V = 0.1V\Earth$ and $p_{\Earth} = 3.3\times10^{-4}$ bar following 
\citet{menou2015}, Fig. \ref{fig:soil} predicts that limit cycles should occur with a frequency 
of about ten cycles per Gyr. Even if we assume values similar to those used by \citet{kadoya2014}, 
with $V = V\Earth$ and $p_{\Earth} = 3.3\times10^{-4}$ bar, this still results in climates 
prone to limit cycling. However, when we select our preferred values of 
$V = V\Earth$ as the present volcanic outgassing rate and $p_{\Earth} = 10^{-2}$ bar as 
the present-day value of pCO$_2$ in soil, then Fig. \ref{fig:soil} predicts that limit cycles should
not occur. Our improvements to the radiative transfer and our consideration of the mass balance of CO$_2$ ice,
which was absent in previous studies, make 
our model more accurate in predicting the frequency of limit cycle events; however, 
our assumptions about $V$ and $p_{\Earth}$ determine when these limit cycles occur.

\begin{figure}
\epsscale{0.8}
\plotone{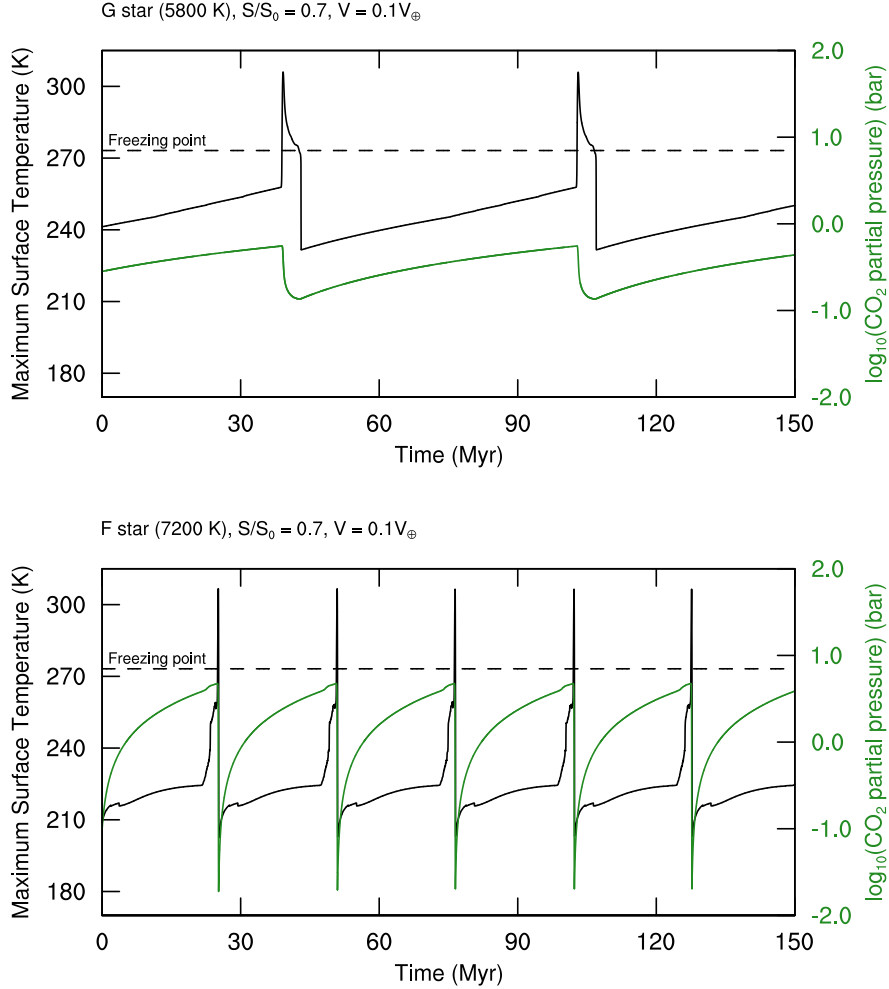}
\caption{Maximum surface temperature (black curve, left axis) and $p$CO$_2$ (green curve, right axis) as a function of 
time for an Earth-like planet with $V=0.1V_{\Earth}$ orbiting a G star (top panel) and F star (bottom panel) at $S/S_0 = 0.7$
Transient periods of warming during which temperatures exceed the freezing point of 
water (dashed line) are followed by extended periods of global glaciation. Earth-like planets that exhibit 
this type of limit cycle cannot maintain permanent surface liquid water and are therefore inhospitable to complex life.
\label{fig:FG}}
\end{figure}

When we use our model to investigate the possibility of limit cycles occurring within the Sun's HZ, 
we find that no such limit cycle boundary exists for our choices of 
volcanic outgassing rate
($V_{\Earth}=70\mbox{ bar/Gyr}$) and soil $p$CO$_2$ ($p_{\Earth}=10^{-2}\mbox{ bar}$).
This model configuration allows an Earth-like planet to deglaciate from a snowball state 
at any point within the conventional HZ due to the accumulation of a dense CO$_2$ atmosphere 
from the carbonate-silicate cycle. 
This result may initially appear inconsistent with Fig. \ref{fig:cycles}, where we predict limit cycles 
for $S/S_0 = 0.43$. However, the climate calculations in Fig. \ref{fig:cycles} are global averages 
from a one-dimensional model \citep{kopparapu2013,kopparapu2014}, in which the surface must be 
either completely ice-free or completely ice-covered. By contrast, our calculations with 
a latitudinal EBM in Fig. \ref{fig:FG} allow polar ice caps to form while still retaining 
ice-free conditions at lower latitudes, which permits stable 
climates to persist across the entire HZ. This illustrates the importance of using latitudinally-resolved 
models for situations where a planet's  surface is partly ice-covered but remains otherwise habitable.
Thus, we still retain a classic picture of the 
HZ, where warm stable climates (even for abiotic planets) are possible all the way out to 
the maximum greenhouse effect. This lack of limit cycling throughout the HZ applies 
to planets orbiting F, G, K, and M stars, as the tuning of our model to soil $p$CO$_2$ results in warm 
stable climates regardless of stellar type.
A decrease in $\beta$ would cause the weathering rate to be even less sensitive 
to changes in $p$CO$_2$, which only expands the 
climatically stable parameter space where limit cycles do not occur.
This suggests that the effect of limit cycling for planets 
orbiting near the outer edge of the HZ may have been overestimated by 
previous studies \citep{kadoya2014,kadoya2015,menou2015}.

We have argued that volcanic outgassing should balance weathering at the soil $p$CO$_2$, 
rather than the atmospheric $p$CO$_2$, which results in no limit cycles for Earth-like 
planets in the HZ. 
However, this result also depends on our assumed volcanic outgassing rate, which we 
have tuned to present-day Earth. \citet{kadoya2014,kadoya2015} have argued convincingly that 
the habitability of an Earth-like planet is highly sensitive to the CO$_2$ degassing rate. 
(See also earlier papers by \citet{tajika2003,tajika2007}.)
Lower rates of volcanism will decrease atmospheric CO$_2$, which will increase 
the susceptibility to limit cycles, 
as we show in Fig. \ref{fig:soil}.
Furthermore, increased fractional land area could accelerate loss of CO$_2$ by 
silicate weathering and make a planet more prone to limit cycles, 
provided that sufficient rainfall is available. Thus, limit cycles could 
occur on planets that have a reduced volcanic outgassing rate or increased 
weathering rate compared to abiotic Earth.

We next consider the effect of volcanic outgassing rate on the occurrence of 
limit cycles in the HZ. 
Beginning with a planet orbiting a G-type star, we decrease the volcanic outgassing 
rate to $V = 0.1V_{\Earth}$, and
we calculate the distance at which limit cycles begin by decreasing the relative 
solar flux $S/S_0$ until cycles start to develop. (Solar flux and orbital distance are 
related by the inverse square law.) This transition occurs at $S/S_0$ = 0.77, 
which corresponds to an orbital distance of about 1.2 AU. At that distance and beyond, the planet's 
climate exhibits warm intervals of $\sim$5 Myr in duration with equatorial temperatures above 
the freezing point of water, separated by extended periods of global glaciation 
lasting $\sim$60 Myr (Fig. \ref{fig:FG}, top panel). During the warm intervals, increased weathering rates lead 
to a decrease in atmospheric CO$_2$, which eventually triggers the subsequent glaciation. 
Because continental weathering ceases entirely during glaciation, atmospheric CO$_2$ accumulates 
from ongoing volcanic outgassing. 

The effects of limit cycling are quite different for hot, blue, F-type stars than for cooler, 
red, K- and M-type stars, largely because of the way in which incident stellar radiation interacts with 
surface water ice. Water ice is highly reflective at visible wavelengths ($<$700 nm), but becomes an 
increasingly efficient absorber at longer, near-infrared wavelengths. Ice-albedo feedback is therefore 
greater for planets around F stars than it is for planets around K and M stars because F stars emit a 
greater percentage of their radiation at visible wavelengths \citep{joshi2012,shields2013}.
For F stars, again assuming $V = 0.1V_{\Earth}$, the limit cycle transition occurs at 
$S/S_0$ = 1.07, which corresponds to an orbital distance of about 0.96 AU. This planet experiences 
warm periods lasting $\sim$1 Myr with prolonged glacial periods of $\sim$20 Myr (Fig. \ref{fig:FG}, bottom panel).
Planets orbiting K- and M-type stars do not experience limit cycles, even at this reduced
volcanic outgassing rate, but instead these systems can maintain stable warm conditions all the way 
out to the maximum greenhouse limit.

\begin{figure}
\epsscale{0.8}
\plotone{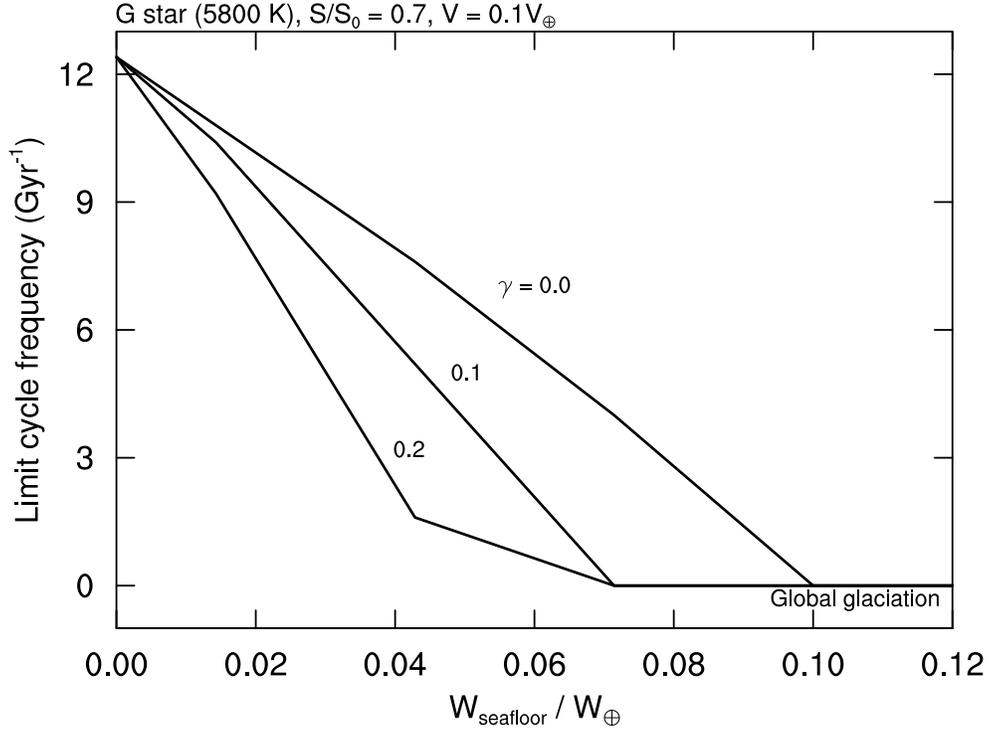}
\caption{Diagram showing the dependence of the limit cycle frequency on seafloor weathering rate for an 
Earth-like planet with $V=0.1V_{\Earth}$ orbiting a G star at $S/S_0 = 0.7$,
with sensitivity to the 
seafloor weathering parameter $\gamma$. Complete decoupling of $p$CO$_2$ from seafloor weathering ($\gamma = 0$) results 
in a steady decline in the frequency of limit cycle warming events as the seafloor weathering rate increases, 
until the uptake of CO$_2$ by the seafloor equals the outgassing rate. When the rate seafloor weathering 
depends on $p$CO$_2$ ($\gamma > 0)$, then the accumulation of CO$_2$ in the atmosphere from outgassing 
drives stronger seafloor weathering and makes the planet more prone to permanent glaciation.
\label{fig:sea}}
\end{figure}

Seafloor weathering \citep{coogan2013} is another potential sink for atmospheric CO$_2$, 
which can make a planet more susceptible to limit cycles or permanent glaciation. 
For a G star planet with $S/S_0$ = 0.7 
and $V = 0.1V_{\Earth}$ that is already experiencing limit cycles, an increase in
the rate of seafloor weathering serves to decrease the frequency in cycles between 
climate states, which results in longer 
extended periods of glaciation between warm episodes (Fig. \ref{fig:sea}). 
The rate of seafloor weathering may also be sensitive to atmospheric 
$p$CO$_2$ through the parameter $\gamma$ in Eq. (\ref{eq:seaweather}), but 
increases in $\gamma$ 
only accentuate the tendency of a planet in a limit cycle toward permanent glaciation. 
For planets beyond the limit cycle region, an increase in seafloor weathering will cause 
the limit cycle region of the HZ to expand, analogous to a decrease 
in volcanic outgassing of the same magnitude.
The rest of our calculations will 
consider sensitivity to the volcanic outgassing rate, $V$, but these results 
apply more broadly to the difference
between volcanic outgassing and seafloor weathering, $V-W$,
that drives long-term changes in $p$CO$_2$ (Eq. (\ref{eq:weathering})).
Strong seafloor weathering could cause a planet to be prone to limit cycles even 
if the volcanic outgassing rate is at or above present-day values.

\begin{figure}
\epsscale{0.8}
\plotone{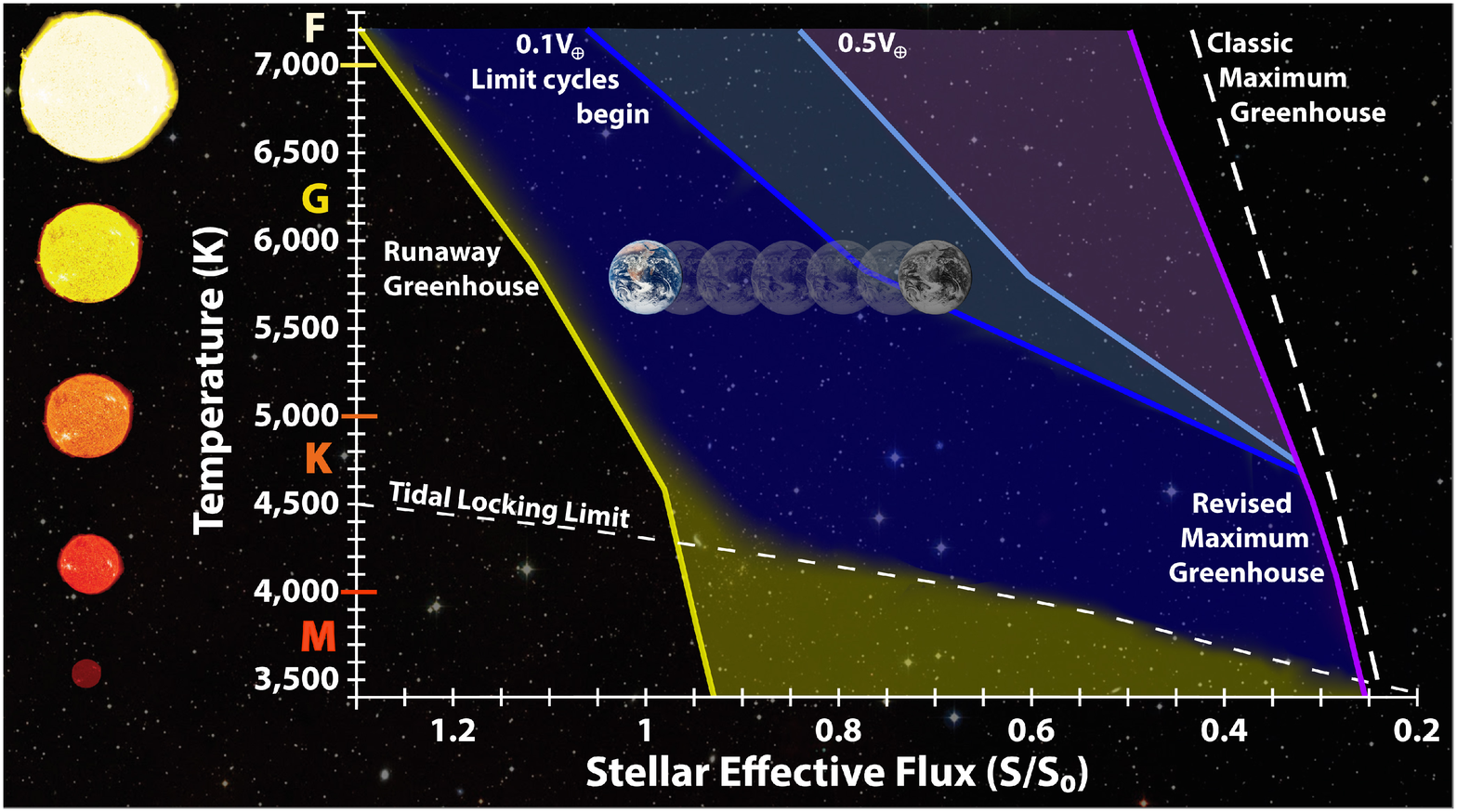}
\caption{Habitable zone boundaries for various stellar types as a function of stellar effective flux,
normalized to that at Earth's orbit today ($S_0$). The `runaway greenhouse' and `classic maximum greenhouse' 
limits (solid blue curves) show the conventional HZ boundaries \citep{kopparapu2013}, 
as corrected to account for recent 3-D studies \citep{leconte2013}. The 
`revised maximum greenhouse' limit (dashed blue curve) shows the flux beyond which an Earth-like planet would be unable to 
deglaciate from an initial snowball state. (The classic maximum greenhouse assumes an ice-free surface, 
which is inconsistent with the limit cycle behavior described here.) Planets beyond this boundary are 
thus uninhabitable at their surfaces. The boundaries labeled `Limit cycles begin' marks the region beyond 
which a stable warm climate cannot be maintained for volcanic outgassing rates in the range 
$0.1V_{\Earth} < V < 0.5V_{\Earth}$.
Modern Earth is shown, along with Earth at 4.5 Gyr ago, 
when the solar flux was $\sim$70 percent of its current value. 
\label{fig:hz}}
\end{figure}

We summarize our findings by performing similar calculations of the limit cycle boundary 
for F, G, K, and M stars with $V = 0.5V_{\Earth}$
and combining these calculations with the results for $V = 0.1V_{\Earth}$
in Fig. \ref{fig:hz}. 
As discussed above, planets with a present-day volcanic outgassing rate avoid limit cycling altogether, 
while a rate of $V = 0.5V_{\Earth}$ or lower can be sufficient to induce cycling for at least 
part of the HZ. Planets orbiting F stars 
exhibit cycling behavior throughout a large region of their HZs and are the most responsive to 
changes in $V$, as a result of their increased sensitivity to ice-albedo feedback.
A planet like 
Earth around a G star should have experienced limit cycles in its past only if volcanic outgassing 
were much lower than today.
Modern Earth avoids this fate, but early Earth may have experienced repeated Snowball Earth episodes 
if volcanic outgassing rates were relatively low \citep{sleep2001,tajika2007}. The rock record is sparse or 
nonexistent during the first half of Earth's history, so such behavior could have occurred but remained undetected. 

By contrast, late K and M star planets can maintain stable climates without cycling until the maximum 
greenhouse limit is reached. Such planets are subject to tidal locking and may show only one face to 
the star, as the Moon does to Earth, which may or may not present a problem for complex life. However, 
planets orbiting these stars may lose their entire water inventory as a consequence of a runaway 
greenhouse during the extended, hot pre-main sequence evolution of the host star 
\citep{ramirez2014,luger2015,tian2015}. This suggests 
that such planets may be entirely uninhabitable unless they begin with an abundant water inventory or 
water is somehow resupplied after the star enters the main sequence. 

Note that the HZ itself is also narrower in Fig. \ref{fig:hz}
than prior estimates because the requirement for a planet to be able to recover from global glaciation 
shifts the outer edge slightly inward in our model. This last result is sensitive to the fraction of 
the planet's surface covered by (lower albedo) bare ice, as opposed to (higher albedo) snow-covered 
ice. In our model the ice is snow-covered at all latitudes, and so its albedo remains relatively 
high, even in the near-infrared. By contrast, \citet{shields2013} found that the outer edge should 
be completely insensitive to surface albedo because its effect would be completely masked by the 
overlying dense CO$_2$-rich atmosphere. However, such dense CO$_2$ atmospheres may not be long-lived 
when carbonate-silicate weathering is included.

\section{Discussion}

The Rare Earth hypothesis \citep{ward1999} suggests that complex life may be uncommon in the universe even if
simple life is widespread. Although many arguments have been raised against this idea \citep{kasting2001},
limit cycling in parts of the conventional HZ, 
for planets with relatively low outgassing rates, presents 
problems for both simple and complex life. Photosynthetic algae and cyanobacteria would go 
extinct on a `hard Snowball' planet with sea-ice thicknesses of a kilometer or more, unless 
illuminated refugia were available. Both types of organisms survived the Neoproterozoic 
Snowball Earth episodes \citep{hoffman1998}, so some types of refugia must have existed. Models with either 
thin ice \citep{pollard2005} or open water \citep{abbot2011} near the equator may provide the explanation. But multicellular 
land life would be highly challenged by this type of climatic behavior which, fortunately, has 
not occurred since the late Precambrian. It follows that animal life, and thus intelligent life, 
may not be able to evolve on planets with low incident stellar flux and a low volcanic outgassing rate, 
even if they are within the conventional HZ.

The presence of a limit cycle boundary depends critically upon the assumed volcanic outgassing rate, 
and planets with a CO$_2$ outgassing rate similar to today may not experience limit cycles at all. 
\citet{kadoya2015} suggest that low outgassing rates may be expected for Earth-like planets orbiting 
old stars, which would make such planets more prone to limit cycles. 
One should exercise caution in accepting this conclusion, as it is based on their assumption 
that the CO$_2$ outgassing rate declines monotonically with time on an Earth-like planet as 
its interior cools. Other authors \citep{holland2009}
have suggested that Earth's CO$_2$ outgassing rate actually {\it increased} with time during the 
first half of its history as the growth of continents allowed greater storage of carbonate 
rocks and greater recycling of CO$_2$ through weathering, carbonate deposition, and sediment subduction.

If the CO$_2$ outgassing 
rate of present-day Earth is anomalously large compared to typical terrestrial planets, then
Earth might be uncommon in its ability to sustain a stable warm climate. 
Conversely, planets more massive than Earth (known as super-Earths) may exhibit higher rates of 
volcanism than Earth today, although the dependence of plate tectonics on planetary mass remains 
unclear \citep{valencia2007,kite2009,korenaga2010,vanheck2011,haghighipour2013}. So, it is at least conceivable that super-Earths in the outer parts of the HZ 
would be better abodes for complex life than would true Earth analogs.

The actual dependence of the silicate weathering rate on $p$CO$_2$ is unknown for abiotic 
planets. We have assumed $\beta = 0.5$,
which matches the behavior of the H$^+$ concentration in rainwater; 
however, the actual exponent could range from 0.25 to 1 \citep{berner1994}. The weathering rate on modern Earth is 
sometimes argued to have zero direct dependence on atmospheric CO$_2$ \citep{berner1983}, because $p$CO$_2$ in soils is 
decoupled from atmospheric $p$CO$_2$ by the presence of vascular plants, which pump up soil CO$_2$ by way of 
root respiration. Plants also generate humic acids which accelerate weathering, again without any 
direct relation to atmospheric $p$CO$_2$. Consequently, \citet{menou2015} argued that the emergence of land life 
on a planet should stabilize its climate against limit cycles. But this inference is incorrect. Land 
plants accelerate weathering \citep{berner1992} by anywhere from a factor of 2-3 \citep{cawley1969} 
to a factor of 10-100 or more \citep{schwartzman1989}
compared to an abiotic environment. (We assume a weathering acceleration factor 
in the range of 2.8-5.5, as dicussed above.)
Lichens, algae, and other microorganisms also secrete acids 
that accelerate weathering \citep{berner1992}, so the emergence of life on a planet in the outer, limit-cycling 
region of the HZ should only help to pull down atmospheric CO$_2$, making the planet even more subject 
to global glaciation. Global glaciation would kill any plants, allowing atmospheric CO$_2$ to again 
accumulate, and so cycling should re-initiate at a rate that would depend on whether the plants 
themselves were able to regenerate. Life (as we know it) would not stabilize a planet's climate 
against limit cycling, but it might create a more complex, biologically mediated form of limit cycling.

What do these arguments imply about the prevalence of animal life and the possible 
evolution of intelligent life? The limit cycle region of the HZ depends upon the assumed, 
and somewhat uncertain, behavior of the volcanic outgassing and seafloor weathering rates for abiotic planets. 
For planets around K and M stars, this does not appear to pose a problem, but for some 
types of stars, the outlook is less optimistic. According to our results in Fig. \ref{fig:hz}, 
the HZ around F stars can be nearly eliminated if the volcanic outgassing rate is $0.1V_{\Earth}$ or less.
F stars also have relatively short main sequence lifetimes, and 
they brighten quickly as they age, which limits the time available for biological evolution. 
Meanwhile, although planets around K and M stars may avoid this problem, they may have other issues 
that could preclude their habitability.
If \citet{ramirez2014}, \citet{luger2015}, and \citet{tian2015} are correct, nearly all planets around late K and M stars should 
experience drastic pre-main-sequence water loss. Early 
K stars and late G stars avoid both of these problems. 
So, there are still many stars that could host planets with complex life. 
But any search for such life should be concentrated on planets around late G- to early K-type stars, 
which are only a subset of the planets that might support simple life. 

\section{Conclusion}

We have calculated the limit cycle boundary of the habitable zone as a function of stellar type 
and CO$_2$ outgassing rate. Earth-like planets with volcanic outgassing rates similar 
to today are able to maintain stable climates across the entire range of the HZ, regardless 
of stellar type. But planets with lower volcanic outgassing rates 
or significant seafloor weathering rates should experience limit cycles, 
with punctuated episodes of warm conditions followed by extended glacial periods. F star planets 
are the most prone to this behavior as a result of increased susceptibility to ice-albedo feedback. 
Planets orbiting late K and M stars avoid limit cycles because of reduced ice-albedo feedback, 
but they may suffer from water loss during their formation. Thus, systems with the greatest 
potential for habitability are those around late G and early K type stars. 

The net outgassing rate of CO$_2$ and partial pressure of $CO_2$ in soil
are key parameters in understanding the habitability of an Earth-like 
planet. If Earth has maintained a net CO$_2$ outgassing rate at or above its current value 
for its entire history, then it may never have been prone to limit cycles at any point 
in time. By extension, if Earth's outgassing rate is typical for other terrestrial planets,
then limit cycles may not pose a problem for habitability at all. 
Likewise, we expect that an abiotic planet would accumulate more atmospheric 
CO$_2$ than its inhabited counterpart, which would lead us to expect that few planets,
if any, should reside in limit cycles. 
However, we 
should expect a diversity of exoplanet environments to exist, and we cannot rule out 
the possibility that Earth's CO$_2$ inventory is atypical.
In the search for habitable worlds, we should at least consider the possibility that some 
Earth-like planets may exhibit lower net outgassing rates 
than today, which could develop limit cycles and preclude the development of complex life.

\acknowledgments
The authors thank Darren Williams for assistance with model development as well as Dorian Abbot, 
Ray Pierrehumbert, Aomawa Shields, and Russell Deitrick for helpful discussions. 
The authors also thank an anonymous reviewer for thoughtful comments which greatly improved the manuscript.
J.H. acknowledges funding from the NASA Habitable Worlds program under award NNX15AQ82G. R.K.K. and 
J.F.K acknowledge funding from NASA Astrobiology Institute's Virtual Planetary Laboratory lead 
team, supported by NASA under cooperative agreement NNH05ZDA001C. 
R.K.K. and J.H. also acknowledge funding from the NASA Habitable Worlds program 
under award NNX16AB61G. This material is based upon 
work supported by the National Science Foundation under Grant No. DGE1255832 to N.E.B. Any 
opinions, findings, and conclusions or recommendations expressed in this material are those 
of the authors and do not necessarily reflect the views of NASA or the National Science Foundation.

\appendix
\section{Polynomial fits to OLR and planetary albedo for F, G, K, and M stars}
We parameterized top of the atmosphere (TOA) albedo, $\alpha$, and the outgoing IR flux, $F_{OLR}$, as
polynomials with the following variables:
surface temperature $T_s$ (K) used as $t=\text{log}_{10}(T_s)$; $\phi = \text{log}_{10}(p$CO$_{2})$, where $p$CO$_{2}$ is the
partial pressure of CO$_2$ (bars);  $\mu = cos(z)$
where $z$ is the zenith angle; and surface albedo $a_{s}$.
The parameterizations were derived by running a 1-D radiative convective (RC) model \citep{kopparapu2013,kopparapu2014} 
over a range of values of
the above parameters with a 1 bar N$_2$ noncondensible background for each stellar type. 
The fits are valid in the range $150$ K$ < T_s < 350$ K, $10^{-5}$ bar $ < p$CO$_{2} < 35$ bar, $0.2 < a_{s} < 1$ and $0 < \mu < 1$.
For planetary albedo, we made separate fits above and below 250 K. 

Solar zenith angle is explicitly calculated at each latitude in the EBM as a function 
of solar declination and solar hour angle, averaging over a complete rotation to 
obtain the insolation-weighted zenith angle \citep{williams1997,cronin2014}. In general, 
we expect that TOA should increase as a function of $z$. The usual configuration of 
our 1-D RC model \citep{kasting1991,kasting1993,kopparapu2013,kopparapu2014} assumes 
that stratospheric temperature $T_{strat}$ is equal to the `skin temperature' of a gray atmosphere 
(i.e., the temperature at an optical depth of zero), given by
\begin{eqnarray} 
 T_{\mathrm{strat}} &=& \frac{1}{2^{1/4}}\left[ \frac{S}{4 \sigma} (1 - \alpha)\right], 
\label{tstrat}
\end{eqnarray} 
where $\sigma$ is the Stefan-Boltzmann constant. 
However, Eq. (\ref{tstrat}) 
is only appropriate for global average conditions (i.e., for $z$ $= 60^{\circ}$) 
and cannot be applied to calculate $T_{strat}$ and $\alpha$ for the range of 
solar zenith angles across all latitude bands represented in our EBM. In order 
to circumvent this problem, we followed \citet{williams1997} and 
parameterized $T_{strat}$ as a function of $z$:
\begin{eqnarray} 
T_{\mathrm{strat}}(Z) &=& T_{strat} (60^{\circ}) \left[ \frac{F_{s}(z)}{F_{s}(60^{\circ})}\right]^{1/4},
\label{tstrat2}
\end{eqnarray} 
where $F_{s}$ is the absorbed fraction of incident solar
flux, which was calculated for a variety of zenith angles between $0^{\circ}$ and $90^{\circ}$ using the
radiative-convective model. The value of $T_\text{strat}$($60^{\circ}$) was obtained using Eq. (\ref{tstrat}) above.
This configuration of our RC model allows the parameterizations of TOA below to accurately decrease 
with $z$ as expected.

\subsection{Parametric expressions for OLR and Planetary Albedo}
The coefficients are provided as downloadable ASCII data tables in online supplementary information. The rows in the data table correspond to the order
of the coefficients in the following expressions. The OLR coefficients are the same for all different stars, as it is
a planet specific property and variation in stellar type does not change it. For planetary albedo coefficients,
each stellar type has two sets of data tables, one for surface temperatures between $150$ K to $ 250$ K and another
for $250$ K to $350$ K.
\begin{align}
\begin{split}
F_{OLR} ={}& A t^{4} + B  t^{3} \phi + C  t^{3} + D t^{2} \phi^{2}+  E  t^{2} \phi + F  t^{2}+ G  t \phi^{3} + H t \phi^{2} + \\
         &   I  t \phi + J  t +  K \phi^{4} + L  \phi^{3} +  M  \phi^{2} + N  \phi +   constant1
\end{split}
\end{align}
\begin{align}
\begin{split}
\alpha (t, a_{s}, \mu, \phi ) ={}& A \mu^{3} +B \mu^{2} a_{s} + C \mu^{2} t + D \mu^{2} \phi + E \mu^{2}+ F \mu a_{s}^{2}+  G \mu a_{s} t+ H \mu a_{s} \phi+ I \mu a_{s}\\
           & +J \mu t^{2}+ K \mu t \phi+ L \mu t + M \mu \phi^{2} + N \mu \phi + O \mu+ P a_{s}^3+ Q a_{s}^{2} t+ R a_{s}^{2} \phi + S a_{s}^{2}\\
          & + T a_{s} t^{2}+  U a_{s} t \phi+ V a_{s} t + W a_{s} \phi^{2}+ X a_{s} \phi+ Y a_{s} + Z t^{3}+ AA t^{2} \phi  + AB t^{2}\\
          &  + AC t \phi^{2}+ AD t \phi+ AE t+ AF \phi^{3}+  AG \phi^{2}+ AH \phi+ constant2
\end{split}
\end{align}
These paramtetric fits provide a rapid means of obtaining OLR and planetary albedo from our 1-D RC climate model calculations.
The error between the climate model data and the parametric fits for OLR does not exceed 2\%, and 
the error distributions of planetary albedo for different stellar spectral types show that a majority of parametric fits have less than 20\% uncertainty.
Error plots for OLR and planetary albedo are provided as online supplementary information.

\end{document}